\documentclass{aa}  

\usepackage{graphicx}
\usepackage{txfonts}
\begin{document}

   \title{High-resolution radio imaging of two luminous quasars beyond redshift 4.5}


   \author{S.~Frey\inst{1}
          \and
          O.~Titov\inst{2}
          \and
          A.~E. Melnikov\inst{3}
          \and
          P.~de~Vicente\inst{4}
          \and
          F.~Shu\inst{5} 
          }

   \institute{Konkoly Observatory, MTA Research Centre for Astronomy and Earth Sciences, Konkoly Thege Mikl\'os \'ut 15-17, H-1121 Budapest, Hungary\\
              \email{frey.sandor@csfk.mta.hu}
         \and
             Geoscience Australia, P.O. Box 378, Canberra, ACT 2601, Australia\\
             \email{oleg.titov@ga.gov.au}
         \and
             Institute of Applied Astronomy, Russian Academy of Sciences, Kutuzov embankment 10, 191187 Sankt-Peterburg, Russia\\
             \email{aem@iaaras.ru}
         \and
             Observatorio de Yebes (IGN), Apartado 148, E-19180 Yebes, Spain\\
             \email{p.devicente@oan.es} 
         \and
             Shanghai Astronomical Observatory, Chinese Academy of Sciences, 80 Nandan Road, Shanghai 200030, China\\ 
             \email{sfc@shao.ac.cn} 
             }

   \date{Received February 5, 2018; accepted July 11, 2018}

 
  \abstract
   {Radio-loud active galactic nuclei in the early Universe are rare. The quasars J0906+6930 at redshift $z$=5.47 and J2102+6015 at $z$=4.57 stand out from the known sample with their compact emission on milliarcsecond (mas) angular scale with high (0.1-Jy level) flux densities measured at GHz radio frequencies. This makes them ideal targets for very long baseline interferometry (VLBI) observations.}
   {By means of VLBI imaging we can reveal the inner radio structure of quasars and model their brightness distribution to better understand the geometry of the jet and the physics of the sources.}
   {We present sensitive high-resolution VLBI images of J0906+6930 and J2102+6015 at two observing frequencies, 2.3 and 8.6~GHz. The data were taken in an astrometric observing programme involving a global five-element radio telescope array. We combined the data from five different epochs from 2017 February to August.}
   {For one of the highest redshift blazars known, J0906+6930, we present the first-ever VLBI image obtained at a frequency below 8~GHz. Based on our images at 2.3 and 8.6~GHz, we confirm that this source has a sharply bent helical inner jet structure within $\sim$3~mas from the core. The quasar J2102+6015 shows an elongated radio structure in the east--west direction within the innermost $\sim$2~mas that can be described with a symmetric three-component brightness distribution model at 8.6~GHz. Because of their non-pointlike mas-scale structure, these sources are not ideal as astrometric reference objects. Our results demonstrate that VLBI observing programmes conducted primarily with astrometric or geodetic goals can be utilized for astrophysical purposes as well.}
   {}


   \maketitle
%

\section{Introduction}
\label{intro}

Active galactic nuclei (AGN) are located in the central region of galaxies. The source of their extreme power, which makes them the most luminous persistent (non-transient) astrophysical objects in the Universe, is accretion of matter onto a supermassive ($\sim 10^6 - 10^{10}$~M$_{\odot}$) black hole \citep[e.g.][]{rees84,kormendy95}. Active galactic nuclei can be detected at multiple wavebands across the entire electromagnetic spectrum, even from extremely large cosmological distances. Their most distant representative known to date is a quasar at redshift $z=7.54$, corresponding to just 5\% of the current age of the Universe \citep{banados18}. A small fraction, about 10\% of AGN, are strong radio sources. The origin of their radio emission is synchrotron radiation produced by charged particles accelerated to relativistic speeds in a strong magnetic field in bipolar jets emanating from the vicinity of the central supermassive black hole \citep[e.g.][]{bridle84,zensus97}. Among the radio AGN, the most distant known at present is at $z=6.21$ \citep{willott10,frey11}. 

Because AGN populate the observable Universe from low to high redshifts, they can be used to study galaxy evolution and test cosmological model predictions. In this respect, objects at very high redshifts are especially valuable. They are also rare because of the limited sensitivity of our instruments. For example, the number of known AGN with measured spectroscopic redshift at $z>4$ is nearly 2600, and only about 170 are known radio sources \citep{perger17}. In comparison, the total number of AGN is now close to a million \citep{flesh15}. 

The compact pc-scale radio structure of AGN jets can be best studied with the technique of very long baseline interferometry \citep[VLBI; see][for a recent review]{middelberg08}. This technique involves coordinated observations of selected celestial objects by radio telescopes separated by large, often intercontinental distances. The technique is applied to reconstruct the brightness distribution of radio sources. In fact, VLBI provides the finest angular resolution currently achievable in astronomy. 

On the other hand, VLBI is also successfully employed for astrometric and global geodetic, geophysical studies \citep[e.g.][]{sovers98}. This technique is unique in regularly and accurately determining the Earth rotation and orientation with respect to a quasi-inertial reference frame realized by the positions of distant, compact radio AGN. The VLBI measurements play an essential role in defining and maintaining the most accurate celestial reference frame \citep{fey15}. Astrometric reference sources preferably have a compact, nearly unresolved appearance in VLBI images, contrary to the objects with rich extended milliarcsecond (mas) scale structures ideal for studying, for example jet physics.  

Only a few radio AGN at extremely high redshift are suitable for detection in astrometric/geodetic VLBI experiments because of the limited baseline sensitivities involving typical 12--40 m geodetic telescopes. Among the total of 30 VLBI-detected extragalactic radio sources at $z>4.5$, only 7 have measured flux densities exceeding $\sim$100~mJy at GHz frequencies \citep{coppejans16}, including the two AGN discussed in this paper.  

The source J0906+6930 was classified as a blazar, i.e. a flat-spectrum radio quasar with one of its relativistic jets approaching and pointing nearly to our line of sight. The redshift of this source is $z=5.47$ \citep{romani04}. The Doppler-boosted jet emission makes J0906+6930 the most radio-luminous quasar and the highest redshift blazar known \citep[e.g.][]{coppejans16,zhang17}. However, if radio flux density measurements below 1~GHz are considered, its overall radio spectrum has a clear turnover at around 6~GHz \citep{coppejans17}. The accurate equatorial coordinates\footnote{International VLBI Service for Geodesy and Astrometry \citep[IVS,][]{schuh12} celestial reference frame, aus2017b solution, \url{ftp://ivs.bkg.bund.de/pub/vlbi/ivsproducts/crf} } of the source are right ascension 09$^{\rm h}$ 06$^{\rm m}$ $30\fs74875$ and declination +69$\degr$ 30$\arcmin$ $30\farcs8363$.

The redshift of the quasar J2102+6015 is $z=4.575$ \citep{sowards04}. The accurate VLBI-determined coordinates are right ascension 21$^{\rm h}$ 02$^{\rm m}$ $40\fs21912$ and declination +60$\degr$ 15$\arcmin$ $09\farcs8364$. Considering its broadband radio spectrum, J2102+6015 also belongs to the gigahertz peaked-spectrum (GPS) class, with an observed-frame turnover frequency $\sim$1~GHz \citep{coppejans17}.

Both sources have been imaged with VLBI at different frequencies in the past, as we discuss it in Sect.~\ref{disc} in the context of our new results. We present new dual-frequency (2.3- and 8.6-GHz) VLBI images of J0906+6930 and J2102+6015 derived from data obtained with a globally distributed array of five radio telescopes located in Russia, Spain, and China. We combined data from a series of five 24 h experiments conducted from 2017 February to August. In Sect.~\ref{obse}, we describe the details of the observations and the data analysis procedure. In Sect.~\ref{resu}, we show the images and the results of modelling the brightness distribution of the sources. In Sect.~\ref{disc}, we compare our results with the information available in the literature and discuss our findings. Finally we give a summary and comment on the potential use of astrometric/geodetic VLBI data for astrophysical purposes in Sect.~\ref{conc}.  

We adopt a standard $\Lambda$CDM cosmological model with parameters $\Omega_{\rm m} = 0.3$, $\Omega_{\Lambda} = 0.7$, and $H_{0} = $70 km\,s$^{-1}$\,Mpc$^{-1}$. In this model, 1~mas angular size corresponds to 6.0~pc and 6.6~pc projected linear size in the rest frame of J0906+6930 and J2102+6015, respectively \citep{wright06}.

\section{Observations and data reduction}
\label{obse}

The quasars J0906+6930 and J2102+6015 were observed along with a large set of bright compact VLBI reference sources in nearly 24 h sessions starting at 19:10 UT each day. The observing log listing the session names, dates, and the participating radio telescopes is given in Table~\ref{log}. The Russian National VLBI Network  Quasar \citep{finkelstein08} consisting of three 32 m diameter radio telescopes at Badary (Bd), Svetloe (Sv), and Zelenchukskaya (Zc) was extended by the 40 m Yebes (Ys) antenna in Spain and the 25 m Sheshan (Sh) radio telescope in China, providing interferometer baselines exceeding 9000~km. The scheduling, observations, and correlation were performed under the auspices of an agreement between the Institute of Applied Astronomy of the Russian Academy of Sciences (IAA RAS) and Geoscience Australia, with the collaboration of the Yebes Observatory in Spain and the Shanghai Astronomical Observatory in China. At each of the five observing epochs, at least four of these stations participated, allowing us to apply amplitude self-calibration \citep{readhead78} during the imaging process.

\begin{table}[t]
\caption{Observing dates and participating VLBI stations}
\label{log}
\begin{tabular}{@{}cccccccc@{}}
\hline
No. & Session & Date & Bd & Sv & Zc & Ys & Sh \\
\hline
1 & RUA012 & 2017 Feb 04--05 & \checkmark & \checkmark & \checkmark & \checkmark &             \\
2 & RUA015 & 2017 Apr 08--09 & \checkmark & \checkmark & \checkmark & \checkmark & \checkmark  \\
3 & RUA016 & 2017 Apr 22--23 & \checkmark & \checkmark & \checkmark & \checkmark & \checkmark  \\
4 & RUA017 & 2017 Aug 19--20 & \checkmark & \checkmark & \checkmark & \checkmark & \checkmark  \\
5 & RUA018 & 2017 Aug 26--27 & \checkmark & \checkmark &            & \checkmark & \checkmark  \\
\hline
\end{tabular}
\end{table}

\begin{figure}
\centering
\includegraphics[bb=15 0 518 495,width=75mm,angle=270,clip]{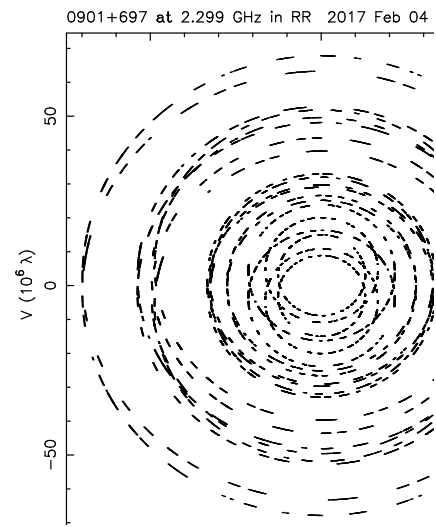}
\caption{The $(u,v)$ coverage of the combined 5-epoch data set for J0906$+$6930 at 2.3~GHz. The points indicate the baseline lengths projected onto a plane perpendicular to the geocentric vector pointing towards the radio source. The $u$ (east--west) and $v$ (north--south) axes are conventionally scaled in the units of million wavelengths ($\lambda$). The nearly circular shape of the $(u,v)$ coverage is caused by the northern hemisphere array observing a circumpolar source at high declination, and the tracks covering full days}
\label{fig:uv}
\end{figure}

Dictated by the primarily astrometric goal of the experiments, the observing schedules were designed in a fashion typical for geodetic/astrometric VLBI observations \citep[e.g.][]{sovers98}. These include consecutive rapid ($\sim$1~min or shorter) scans on multiple radio reference sources located at widely separated elevations, to achieve sufficient spatial and temporal sampling of the observables that are the baseline-based group delays. These scans were interleaved with longer (6--16~min) sections spent on J0906+6930 and J2102+6015. The accurate astrometric positions of the two high-redshift target sources, and their possible changes over a longer period of time will be analysed later and published elsewhere (O. Titov et al., in preparation). The total observing time spent on J0906+6930 and J2102+6015 was about 18 h and 26.5 h, respectively, during the five experiments (Table~\ref{log}).

The VLBI radio telescopes observed with geodetic receivers sensitive at two central frequencies, 2.3 GHz (S band) and 8.6 GHz (X band). A total of 16 intermediate frequency channels (IFs) were used in right-hand circular polarization. The S and X bands were covered by 6 and 10 IFs, respectively. Each IF was divided into 32 spectral channels of 500~kHz width (except for epoch 5, when 8 times more, but also 8 times narrower channels were used). Therefore the total bandwidth was 96~MHz at 2.3~GHz observing frequency and 160~MHz at 8.6~GHz. 
The recorded data were correlated at the IAA RAS in Saint Petersburg with the DiFX software correlator \citep{deller11}. The DiFX shares the hybrid HPC cluster in the IAA Correlator Centre \citep{surkis18} along with the RASFX software correlator \citep{surkis17}. The data were correlated with 0.5~s integration time and the output was converted to FITS-IDI format.

We used the US National Radio Astronomy Observatory (NRAO) Astronomical Image Processing System\footnote{\url{http://www.aips.nrao.edu}} ({\sc aips}) for data calibration in a standard way \citep[e.g.][]{diamond95}. After loading and inspecting the interferometer visibility data supplied by the correlator, the S-band and X-band parts were copied into separate files for independent calibration and analysis. The amplitudes were calibrated using the known antenna gain curves and the system temperatures regularly measured at the antenna sites during the experiments. In cases in which the system temperature measurements appeared unreliable (e.g. for S-band observations at epoch 1), nominal values were used instead. Fringe-fitting \citep{schwab83} was performed with 6 min and 3 min solution intervals at S and X bands, respectively, for all observed radio sources. The delay and delay-rate solutions as a function of time were inspected and a few outliers removed. The solutions were then applied to the visibility data of five selected calibrator sources, J0607+6720, J0726+7911, J1159+2914, J1800+7828, and J0841+7053 (missing from the schedule and replaced by J2321+3204 at the fifth epoch). These data were exported to the Caltech {\sc difmap} software \citep{shepherd94} for imaging. Hybrid mapping cycles with {\sc clean} iterations \citep{hogbom74} and phase-only self-calibration \citep{readhead78} resulted in sufficiently good source models such that overall antenna gain factors could be determined from the data when at least four antennas were present. 

Calibrating the amplitudes in VLBI to the absolute scale is problematic because flux density standards have an extended structure that would be mostly resolved out on the long baselines. It is customary to assume 5--10\% uncertainty of the amplitude calibration. In practice, one has to rely on the system temperature measurements. However, the internal consistency of the amplitudes can be improved if antenna-based gain corrections are determined using scans on bright compact sources scheduled in the experiment. Such scans were available and we selected the above five different calibrators for this purpose. The method is not sensitive to the flux density variability of the calibrator sources. We independently determined the median correction factors in {\sc difmap} for each epoch and each observing frequency and then used them to multiply the antenna gains in {\sc aips}.
Even though the network consisted of four or five stations (Table~\ref{log}), and occasionally nominal system temperatures had to be applied in the absence of measured values, the median antenna gain corrections were remarkably consistent between the different epochs at both observing bands. The standard deviations were between 4\% (average gain correction $0.94\pm0.04$ for Sh at X band) and 18\% ($1.09\pm0.20$ for Sv at S band).    

For our target sources, J0906+6930 and J2102+6015, the calibrated VLBI visibility data with the adjusted antenna gains were transferred to {\sc difmap} for imaging. Hybrid mapping was performed for each individual data set, first with phase-only, then with amplitude and phase self-calibration. We also fitted simple Gaussian brightness distribution model components \citep{pearson95} directly to the self-calibrated visibility data in {\sc difmap}. This facilitates quantitative characterization of the source structure. After producing source images and brightness distribution models at five epochs from 2017 February to August (Table~\ref{log}), we concluded that the flux density of neither J0906+6930, nor J2102+6015 appeared variable.
The standard deviations of the peak brightness values of individual-epoch images restored with the same circular Gaussian beam fell between 4\% and 13\%. Similar epoch-to-epoch variations were seen in the zero-spacing visibility amplitudes and the fitted Gaussian model component flux densities, consistent with the expected amplitude calibration uncertainties. For J0906+6930, an independent confirmation of the quiescence could also be obtained from the 15 GHz flux density monitoring programme at the 40 m telescope of the Owens Valley Radio Observatory \citep[OVRO;][]{richards11}. According to the measurements available\footnote{\url{http://www.astro.caltech.edu/ovroblazars}} at 17 epochs between 2017 February and August, the mean 15 GHz total flux density of J0906+6930 was 81.4~mJy with a standard deviation of $\sim$5\%.

Therefore we concatenated all five calibrated visibility data sets for two sources and two observing frequencies, and repeated the imaging and model-fitting procedure described above. The resulting nearly circular and uniform $(u,v)$ coverage is illustrated by one example shown in Fig.~\ref{fig:uv}. By combining the data taken at different epochs, we were able to decrease the image noise level by almost a factor of 2 compared to the results from individual epochs.

\section{Results}
\label{resu}

\begin{figure*}
\centering
\includegraphics[bb=0 72 885 530,width=160mm,clip]{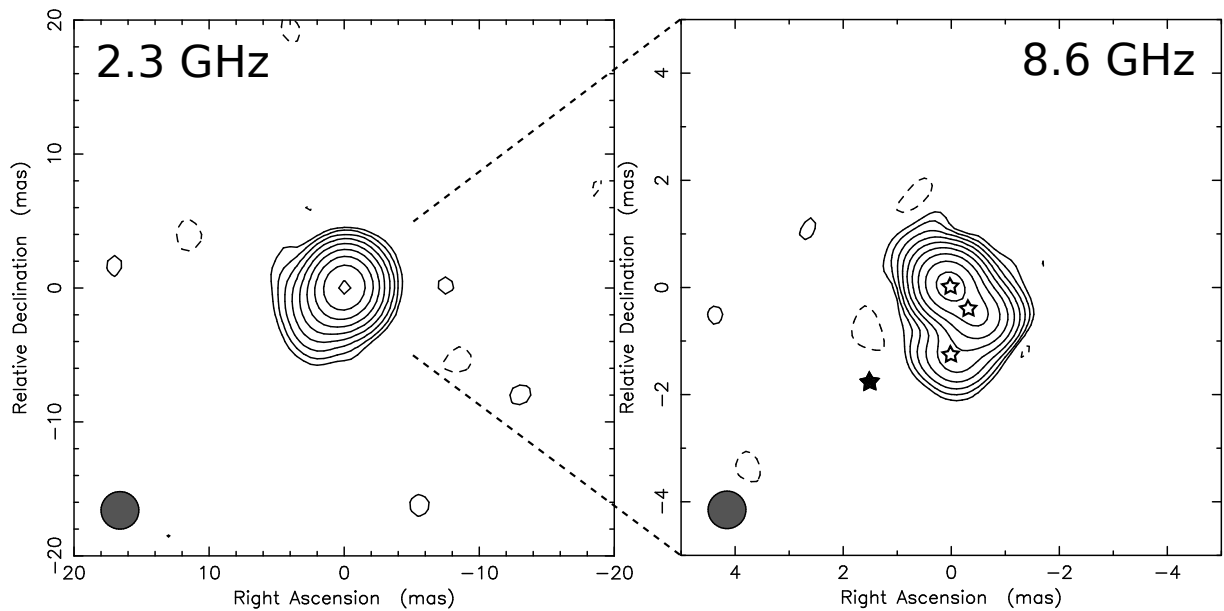}
\caption{Naturally weighted VLBI images of J0906+6930. {\em Left:} at 2.3~GHz. The peak brightness is 95.6~mJy\,beam$^{-1}$. The lowest contours are at $\pm 0.35$~mJy\,beam$^{-1}$ ($\sim$3$\sigma$), the positive contour levels drawn with solid lines increase by a factor of 2. The image is restored with a circular Gaussian beam of 2.8~mas (FWHM) as indicated in the lower left corner. {\em Right:} at 8.6~GHz. The peak brightness is 70.9~mJy\,beam$^{-1}$. The lowest contours are at $\pm 0.2$~mJy\,beam$^{-1}$ ($\sim$3$\sigma$), the positive contour levels increase by a factor of 2. The image is restored with a circular Gaussian beam of 0.7~mas (FWHM). Open star symbols indicate the locations of fitted Gaussian brightness distribution model components at 8.6 GHz. The filled symbol shows the position of the jet component fitted to the 2.3 GHz data (see Table~\ref{model})}
\label{fig:J09}
\end{figure*}

\begin{figure*}
\centering
\includegraphics[bb=0 72 885 530,width=160mm,clip]{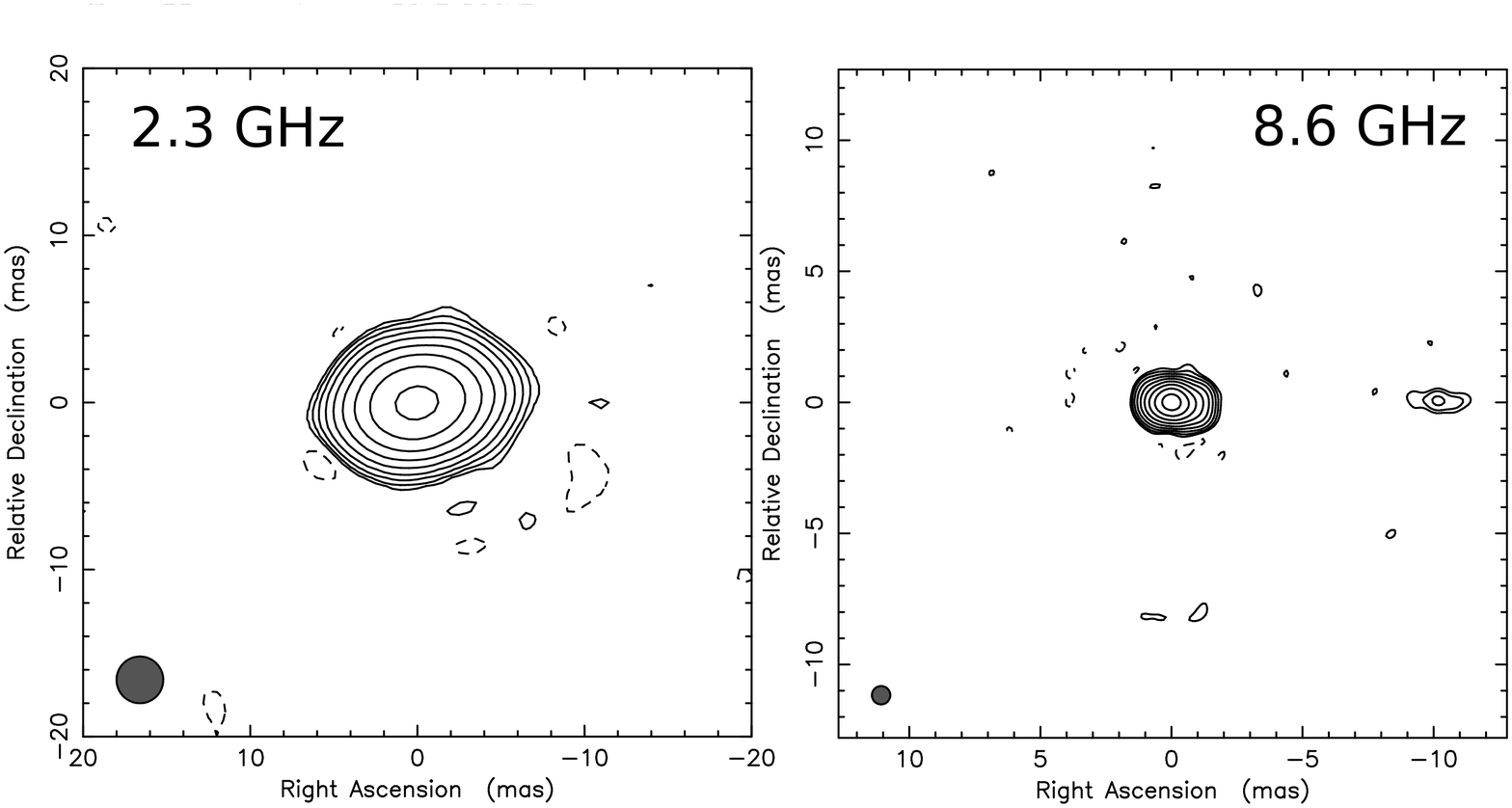}
\caption{Naturally weighted VLBI images of J2102+6015. {\em Left:} at 2.3~GHz. The peak brightness is 123.3~mJy\,beam$^{-1}$. The lowest contours are at $\pm 0.4$~mJy\,beam$^{-1}$ ($\sim$3$\sigma$); the positive contour levels drawn with solid lines increase by a factor of 2. The image is restored with a circular Gaussian beam of 2.8~mas (FWHM) as indicated in the lower left corner. {\em Right:} at 8.6~GHz. The peak brightness is 82.4~mJy\,beam$^{-1}$. The lowest contours are at $\pm 0.22$~mJy\,beam$^{-1}$ ($\sim$3$\sigma$), the positive contour levels increase by a factor of 2. The image is restored with a circular Gaussian beam of 0.7~mas (FWHM)}
\label{fig:J21}
\end{figure*}

We present our best-quality images using the combined five-epoch data for J0906+6930 (Fig.~\ref{fig:J09}) and J2102+6015 (Fig.~\ref{fig:J21}) at 2.3 and 8.6~GHz. The cosmological time dilation makes any structural change in the jet by a factor of $(1+z)$ slower in the observer's frame than in the rest frame of the quasar. Because of the high ($z>4.5$) redshift of both sources, jet component proper motions must appear very slow and thus are negligible during the period covered by our observations. In fact a decade-long time span is needed to reliably detect  jet component proper motions in extremely distant radio quasars with VLBI imaging \citep[e.g.][]{frey15,perger18}. Indeed, investigating closure phases \citep{jennison58} in {\sc difmap} for radio telescope triangles indicates that the structures are consistent for both sources at each individual observing epoch. We therefore believe that the combination of VLBI data obtained in our five different experiments (Table~\ref{log}) is justified and we associate 15\% uncertainty with the derived peak brightness and flux density values to account for both the amplitude calibration errors and the uncertainty resulting from some mild source variability. 

When fitting model components to the visibility data, we kept the number of parameters at the minimum. We used elliptical Gaussians for the brightest central core components to better describe the innermost, barely resolved jet structure, and circular Gaussians for the well-resolved outer jet components. The exception was J2102+6015 at 8.6~GHz where the axial ratio of the core elliptical converged to unity so we replaced it with a circular Gaussian model component. The parameters obtained are listed in Table~\ref{model}; the uncertainties are estimated following \citet{lee08}. The central core components are denoted with 0 and further components are numbered from 1 to 3 and are ordered by increasing distance from the centre. The fitted size of the outermost weak component 3 of J2102+6015 at 8.6~GHz became smaller than the minimum resolvable angular size \citep{kovalev05,martividal12} with this interferometer array, therefore we give that latter value as an upper limit. 

The brightness temperatures of the core components given in Table~\ref{model} are calculated as
\begin{equation}
 T_{\rm b} = 1.22\times 10^{12} (1+z) \frac{S}{\theta_{\rm maj} \theta_{\rm min}\nu ^{2}} \,\, {\rm K}, 
\label{equ-tb}
\end{equation}
where $S$ is the flux density in Jy, $\theta_{\rm maj}$ and $\theta_{\rm min}$ are the elliptical Gaussian component major and minor axes, respectively, in mas, and $\nu$ is the observing frequency in GHz. 
To calculate the monochromatic radio powers in Table~\ref{model}, we applied the formula
\begin{equation}
 P = 4 \pi D^2_{\rm L} S (1+z)^{-1-\alpha}, 
\label{equ-p}
\end{equation}
where $D_{\rm L}$ is the luminosity distance. To obtain values at the same frequencies in the rest frame of the sources, we used a two-point spectral index $\alpha$ (defined as $S \propto \nu^{\alpha}$) calculated from our VLBI measurements. Because of the different angular resolutions at the two observing bands, we associated the 2.3 GHz core with the innermost three fitted Gaussian componets at 8.6 GHz in the cases of both sources. The spectral index is thus $+0.06$ for J0906+6930, and $-0.50$ for J2102+6015.

\begin{table*}
\centering
\small
\caption{Fitted brightness distribution model parameters, calculated core brightness temperatures, and monochromatic radio powers for J0906+6930 and J2102+6015 from the combined 5-epoch VLBI data at 2.3 and 8.6~GHz}
\label{model}
\begin{tabular}{@{}ccccccccccc@{}}
\hline
Source     & $\nu$ & Component & $S$   & $r$ & $\phi$ & $\theta_{\rm maj}$ & $\theta_{\rm min}$ & $\psi$ & $T_{\rm b}$ & $P$ \\
           & GHz   &           & mJy   & mas &  \degr & mas                & mas             &  \degr  & $10^{11}$\,K & $10^{27}$ W\,Hz$^{-1}$\\
\hline
J0906+6930 & 2.3   & 0         & 107.5$\pm$16.1 & 0              & 0              & 1.42$\pm$0.05 & 0.67$\pm$0.03 & 171  &  1.76$\pm$0.28 & 4.8$\pm$0.7 \\
           &       & 1         &  10.8$\pm$1.6  & 2.46$\pm$0.08  & 137$\pm$2      & 1.93$\pm$0.16 & ...           & ...  & ... & ... \\
           & 8.6   & 0         &  79.9$\pm$12.0 & 0              & 0              & 0.38$\pm$0.01 & 0.08$\pm$0.01 & 43   &  2.81$\pm$0.45 & 3.6$\pm$0.5 \\
           &       & 1         &  29.1$\pm$4.4  & 0.74$\pm$0.01  & 224$\pm$1      & 0.14$\pm$0.02 & ...           & ...  & ... & ... \\
           &       & 2         &   6.6$\pm$1.0  & 1.26$\pm$0.02  & 184$\pm$1      & 0.36$\pm$0.05 & ...           & ...  & ... & ... \\
J2102+6015 & 2.3   & 0         & 304.7$\pm$45.7 & 0              & 0              & 4.26$\pm$0.18 & 2.56$\pm$0.08 & 101  &  0.36$\pm$0.06 & 27.3$\pm$4.1 \\
           & 8.6   & 0         &  95.8$\pm$14.4 & 0              & 0              & 0.44$\pm$0.02 & ...           & ...  &  0.46$\pm$0.07 & 8.6$\pm$1.3 \\
           &       & 1         &  35.5$\pm$5.3  & 0.62$\pm$0.02  & 260$\pm$2      & 0.75$\pm$0.05 & ...           & ...  & ... & ... \\
           &       & 2         &  26.0$\pm$3.9  & 0.53$\pm$0.02  &  95$\pm$2      & 0.58$\pm$0.03 & ...           & ...  & ... & ... \\
           &       & 3         &   1.1$\pm$0.2  &10.13$\pm$0.03  & 270$\pm$1      & $<0.2$        & ...           & ...  & ... & ... \\
\hline
\end{tabular}
\\
Notes: Col.~2 -- observing frequency, Col.~4 -- flux density, Col.~5 -- radius from the central component, Col.~6 -- position angle, Col.~7 -- elliptical Gaussian component major axis or circular Gaussian component diameter (FWHM), Col.~8 -- elliptical Gaussian component minor axis (FWHM), Col.~9 -- elliptical Gaussian component major axis position angle; position angles are measured from north through east, Col.~10 -- brightness temperature, Col.~11 -- monochromatic radio power at the given frequency in the rest frame of the source.
\end{table*}

\section{Discussion}
\label{disc}

\subsection{J0906+6930}

The first VLBI images of this source were made by \citet{romani04} at 15 and 43~GHz using the NRAO Very Long Baseline Array (VLBA). At 15~GHz, with their restoring beam of 1.55~mas $\times$ 0.47~mas full width at half maximum (FWHM) elongated in the north--south direction, they detected a core--jet structure with a bright compact central component and a weak jet feature within 1~mas at a position angle 225\degr. This location perfectly matches our model fit results obtained at 8.6~GHz (Table~\ref{model}). Apart from the core, \citet{romani04} also marginally detected the weak jet component at $\sim 3 \sigma$ level at 43~GHz.

The 15~GHz VLBA snapshot observations of \citet{romani04} were not sufficiently sensitive to reveal more details of the jet in J0906+6930. However, \citet{zhang17} re-analysed the 15 GHz data used by \citet{romani04}, concatenated with archival data from other two VLBA experiments conducted in 2005. Their combined image restored with a circular beam (0.6~mas FWHM) revealed a secondary jet feature about 1.5~mas south of the core. \citet{zhang17} pointed out that the existence of the southern jet component would imply a sharp bending of the jet from the south-west to the south. Our 8.6 GHz VLBI observations are well suited to confirm this jet bending because the observing frequency is lower and thus the steep-spectrum jet components are expected to show up more prominently. Moreover, our long projected baselines resulted in a small restoring beam size (0.7~mas, Fig.~\ref{fig:J09}) at 8.6~GHz, comparable to that of the combined 15 GHz VLBA image \citep[Fig.~1c of][]{zhang17}. Indeed, our 8.6 GHz image (Fig.~\ref{fig:J09}) and the modelfit results (Table~\ref{model}) clearly confirm the existence of a bent jet structure in J0906+6930.

A detailed account of all available VLBI observations of J0906+6930 has recently been given by \citet{zhang17}. According to their list, the lowest frequency VLBI image of the source to date was obtained at 8.4~GHz from a VLBA experiment in 2011 June\footnote{Project code BC196, unpublished; available from \url{http://astrogeo.org}}. The short 7 min observing time was sufficient to reveal the compact core only. Our 8.6 GHz image (Fig.~\ref{fig:J09}, right) has a higher resolution and better sensitivity and shows the bent jet structure within $\sim$1.5~mas from the centre. 

Our image at 2.3~GHz (Fig.~\ref{fig:J09}, left) is the lowest frequency VLBI image of J0906+6930 published to date. The overall picture shows a source extended to the south-east that can be modelled with two components (Table~\ref{model}). The extended jet component is located at about 2.5~mas from the centre, at a position angle 137\degr. Therefore the jet bending first noticed by \citet{zhang17} and confirmed by our 8.6 GHz data seems to continue further away from the core. For easy comparison with the 8.6 GHz images, we marked the position of this component in the 8.6 GHz image (Fig.~\ref{fig:J09}, right). We note that the source has no significant extended emission detected on larger, arcsec angular scales \citep{zhang17}.

Based on our dual-frequency VLBI results, the blazar J0906+6930 seems to have a helical mas-scale jet, which we see in projection in the sky \citep[e.g.][]{gower82,conway93}. This is a commonly observed phenomenon in radio AGN and is often attributed to the precession of the nozzle of the relativistic jet pointing close to our line of sight \citep[e.g.][]{lister03,caproni04,molina14,kun14}, but may be caused by a helically twisted pressure maximum in the jet flow \citep{perucho12}, or a magnetically-driven rotating helix \citep{cohen17}. At very high redshifts, precessing jet motion has recently been inferred from modelling a long series of VLBI imaging data for the luminous blazar J0017+8135 at $z$=3.37 \citep{rozgonyi16}. In the case of J0906+6930, a similar kinematic analysis would also require VLBI monitoring data on decadal timescales.

An alternative explanation of the bent jet structure in J0906+6930 is that the jet is deflected by the interaction with the dense ambient gas \citep[e.g.][]{akujor91,fejes92}. It is especially likely to occur in young radio sources in which the jets are expanding within the clumpy interstellar matter \citep[e.g.][]{saxton05,jeyakumar09}.

Our measured brightness temperatures of the core, $T_{\rm b} \approx 2 \times 10^{11}$\,K (Table~\ref{model}) are very close to those derived by \citet{zhang17} from VLBI data taken at multiple frequencies and epochs. This value indicates that the emission of the inner radio jet is Doppler-boosted as expected for blazars. If we assume an intrinsic brightness temperature $T_{\rm b,int} \approx 3 \times 10^{10}$ found by \citet{homan06} for pc-scale jets in a sample of AGN in quiescent state, the Doppler factor is $\delta=T_{\rm b}/T_{\rm b,int} \approx 7$. Future detection of jet component proper motion with VLBI would allow the estimation of the Lorentz factor and the jet inclination angle with respect to the line of sight for J0906+6930.   

The monochomatic radio powers we derived at the rest-frame frequencies of 2.3 and 8.6~GHz are $P \approx 4 \times 10^{27}$\,W\,Hz$^{-1}$ (Table~\ref{model}). These are typical values for other $z>4.5$ radio AGN \citep{coppejans16}, but are significantly, almost an order of magnitude lower than derived by \citet{coppejans16} for the same source, J0906+6930, based on 15 GHz and 43 GHz VLBI measurements. For the explanation, we recall the peaked radio spectral shape of the source \citep{coppejans17}. The turnover frequency $\nu_{\rm t} \approx 6$\,GHz \citep{coppejans17} falls between our two observing frequencies. Hence we observed J0906+6930 at around it spectral peak. Indeed, we found a flat (slightly inverted) spectrum with a two-point spectral index $\alpha=+0.06$. On the other hand, the higher frequencies of 15 and 43~GHz are in the optically thin part of the spectrum characterized by a spectral index close to $-1$ \citep{coppejans16,coppejans17,zhang17}. The $K$-correction term $(1+z)^{-1-\alpha}$ in Eq.~\ref{equ-p} strongly depends on the spectral index, leading to a much higher estimate of the radio power for a source with steep spectrum.

\subsection{J2102+6015}

This quasar has been imaged at 2.3 and at 8.3 and 8.6~GHz in the framework of the first and sixth VLBA Calibrator Surveys\footnote{\url{http://www.vlba.nrao.edu/astro/calib/} } \citep[VCS;][]{beasley02,petrov08}. These snapshot experiments were made in 1994 August and 2006 December. Despite its prominence in flux density and luminosity, no VLBI imaging observations of this high-resdhift quasar have been published at any other waveband to date \citep[cf.][]{coppejans16}. Our new images were made at the same S and X frequency bands as used for VCS, but have higher angular resolution, more uniform $(u,v)$ coverage, and better sensitivity. 

\citet{coppejans17} found that the overall radio spectrum of J2102+6015 is peaked at $\sim$1~GHz (corresponding to $\sim$5.7~GHz in the source rest frame). High-frequency peakers are believed to be genuinely young radio sources. Because of its linear extent of $\sim$70~pc corresponding to the largest angular distance between its 8.6-GHz VLBI components (Table~\ref{model}), J2102+6015 fits well to the turnover frequency--linear size relation derived by \citet{orienti14}.

At 2.3~GHz (Fig.~\ref{fig:J21}, left), we see a single slightly resolved component elongated close to the east--west direction. It is consistent with the shape and orientation of the fitted elliptical Gaussian model component (Table~\ref{model}). The 2.3 GHz flux density of this component, about 300~mJy, is identical with the values derived from the 1994 and 2006 VCS data within the errors \citep[see Table~5 in][]{coppejans16}. 

In the 8.6 GHz image (Fig.~\ref{fig:J21}, right), the east--west elongation remains the most distinctive characteristics. The model fit decomposes the central structure within $\sim1$~mas radius from the brightness peak into three different circular Gaussian components of comparable size (Table~\ref{model}). There are two nearly symmetrical components on both sides of the brightest, central component. Since the absolute astrometric information on the location of the brightness peak is lost with fringe-fitting in VLBI, we are unable to associate any of these 8.6 GHz components with the 2.3 GHz component at the brightness peak. This could unambiguously be done with phase-referencing observations \citep[e.g.][]{beasley95} in the future. Also, nearly simultaneous multi-frequency VLBI observations providing at least as high angular resolution as our 8.6 GHz data would be required to  resolve further the complex  mas-scale radio structure of J2102+6015 and to obtain spectral information to possibly pinpoint a flat-spectrum core.

The brightness temperatures at both frequencies are moderate, around $2 \times 10^{10}$\,K (Table~\ref{model}), suggesting no considerable Doppler boosting in the jet. The overall radio spectrum peaking at around 1~GHz \citep{coppejans17}, the relatively steep spectrum ($\alpha=-0.50$) between our two observing frequencies that are beyond the spectral peak, the high radio power ($P>10^{27}$\,W\,Hz$^{-1}$, Table~\ref{model}), and the compact ($\ll 1$\,kpc) radio structure are all consistent with the classification of J2102+6015 as a GPS source \citep[e.g.][]{odea98}.   

The high-resolution VLBI images of J2102+6015 bear some similarity with those of PKS 1413+135 (J1415+1320), a low-redshift ($z$=0.247) source with two-sided pc-scale structure \citep{perlman96}. J1415+1320 is believed to be a young radio galaxy with an inner structure reminiscent of the sub-kpc size compact symmetric objects \citep[CSOs; e.g.][]{an12}. 

A weak 1.1~mJy component appears at 10.1~mas west of the brightness peak at 8.6~GHz (Fig.~\ref{fig:J21}, Table~\ref{model}). This emission is not seen at 2.3~GHz, even though our image noise level at the lower frequency would allow a compact $\sim$1-mJy component to be detected at $\sim$9$\sigma$ level. The non-detection at 2.3~GHz suggests that its spectrum is inverted, i.e. the component is weaker at the lower frequency. We note that a component at this position is also present in the 8.6 GHz VCS image made in 2006 \citep{petrov08}, but remains a hint in their 2.3 GHz image because of the large restoring beam. We re-analysed the archival data and concluded that in 2006 this component had about five times higher flux density, which may also explain why it was difficult to detect in 2017.

\subsection{Suitability as astrometric reference sources}

According to our imaging VLBI data, both J0906+6930 and J2102+6015 are resolved and show radio structures extended to scales of several mas at the frequency bands routinely used for astrometric and geodetic VLBI observations. Such structures may have adverse effect on the accuracy of parameter estimates and could limit the overall precision of the reference frame \citep[e.g.][]{charlot90,xu17}. One way of characterizing the astrometric quality of the sources is calculating their structure index \citep{fey97} and its modified value for a continuous scale \citep[SI$_{\rm c}$; ][]{fey15}. Based on our X-band data, the SI$_{\rm c}$ values for J0906+6930 and J2102+6015 are 3.0 and 2.9, respectively. Because these values are close to the peak of SI$_{\rm c}$ distribution, these are typical for the sources in the second realization of the International Celestial Reference Frame (ICRF2) but are at the high end of acceptable for defining sources \citep{fey15}. Moreover, these high-redshift sources with correlated flux densities of $\la$100~mJy on long baselines are too weak for regular VLBI observations in the global geodetic network and it is difficult to monitor their positional stability over a long time. Thus the potential of these quasars as suitable astrometric reference objects is limited in future realizations of the ICRF \citep{fey15} or in the most precise alignment of the radio and optical reference frames \citep[e.g.][]{bourda08}. 

\section{Conclusions}
\label{conc}

We reported on a series of VLBI observations conducted at five epochs between 2017 February and August. In each day-long experiment, four or five globally distributed radio telescopes participated, providing interferometer baselines longer than 9000~km. The original purpose of the experiments was studying the accurate astrometric positions of two radio AGN at extremely high redshifts ($z>4.5$), but the VLBI visibility data could also be utilized for high-resolution, high-sensitivity imaging of the sources at 2.3 and 8.6~GHz frequencies. In the absence of significant variability during the period covered by the experiments, we were able to combine the five data sets to improve the image quality. 

One of the highest redshift blazars, J0906+6930 ($z=5.47$), is a well-studied object with VLBI at multiple frequencies \citep{romani04,zhang17}. However, the new observations at 2.3 GHz and 8.6~GHz presented in this paper could significantly advance our knowledge about the source. At 8.6~GHz, we clearly confirmed that the Doppler-boosted inner jet undergoes a sharp bending, as speculated recently by \citet{zhang17} from the analysis of archival 15 GHz VLBA data. Moreover, our 2.3 GHz image indicates that the jet bending continues until at least 2.5~mas distance from the centre. This is the lowest frequency VLBI image of J0906+6930 made to date. Future studies of the newly discovered, apparently helical jet would benefit from even lower frequency imaging since it would be more sensitive to the structure possibly extending to $\sim$10 mas scales. The source with its prominent jet structure almost unique at $z>5$ also has a high potential for studies of jet kinematics. To this end, monitoring over a decade-long time interval is needed because any structural change is expected to be slow in the observer's frame owing to the cosmological time dilation.

As a quasar with one of the highest measured radio powers among currently known AGN at $z>4.5$ \citep[$P>10^{28}$~W\,Hz$^{-1}$;][see also Table~\ref{model}]{coppejans16}, J2102+6015 ($z=4.57$) would deserve more attention. The 8.6 GHz VLBI image presented in this paper has the highest resolution obtained for this GPS source to date. We found that its mas-scale radio structure is elongated in the east--west direction. The rather symmetric central $\sim$2-mas section is resolved into three components for the first time. The accurate characterization of this structure and the secure identification of the radio core in J2102+6015 would require follow-up multi-frequency VLBI observations at 15~GHz and higher frequencies.

Because of their compact radio structure resolved on mas angular scales and correlated flux densites below $\sim$100~mJy, J0906+6930 and J2102+6015 are not ideal as astrometric reference objects, which preferably have point-like structure unresolved on VLBI baselines.

Our results demonstrate the potential in the dual use of S/X-band geodetic/astrometric VLBI experiments. Under certain circumstances, for example for densely sampled monitoring of bright reference objects for a long period of time \citep[e.g.][]{britzen94}, or as in our case, studying selected targets sufficiently bright for fringe-fitting in more detail, the data can be utilized for astrophysical purposes as well. This is made more convenient by the ability of modern correlators operating at geodetic VLBI processing centres to produce outputs containing interferometric visibility data. Proper amplitude calibration requires that the system temperatures are regularly measured at the participating radio telescopes during the observing sessions.

\begin{acknowledgements}
SF thanks for the support received from the Hungarian Research, Development and Innovation Office (OTKA NN110333), and from the China--Hungary Collaboration and Exchange Programme by the International Cooperation Bureau of the Chinese Academy of Sciences. This research has made use of data from the OVRO 40 m monitoring programme \citep{richards11}, which is supported in part by NASA grants NNX08AW31G, NNX11A043G, and NNX14AQ89G and NSF grants AST-0808050 and AST-1109911. The Astrogeo Center Database of brightness distributions, correlated flux densities, and images of compact radio sources produced with VLBI is maintained by L. Petrov. We thank Xuan He for calculating the structure indices.
\end{acknowledgements}

\end{document}